\documentclass[%
 reprint,
 amsmath,amssymb,
 aps,
]{revtex4-2}

\usepackage{graphicx}% Include figure files
\usepackage{dcolumn}% Align table columns on decimal point
\usepackage{bm}% bold math
\usepackage[utf8]{inputenc}
\usepackage{physics}
\begin{document}

\preprint{APS/123-QED}

\title{Measurement of ultrashort bi-photon correlation times with an integrated two-colour broadband SU(1,1)-interferometer}
\author{F. Roeder, R. Pollmann, M. Stefszky, M. Santandrea, K.-H. Luo, V. Quiring, R. Ricken, C. Eigner, B. Brecht, C. Silberhorn}
\affiliation{Paderborn University, Department of Physics, Integrated Quantum Optics, Institute for Photonic Quantum Systems (PhoQS), Warburger Straße 100, 33098 Paderborn, Germany}

\date{\today}% It is always \today, today,

\begin{abstract}
The bi-photon correlation time, a measure for the conditional uncertainty in the temporal arrival of two photons from a photon pair source, is a key performance identifier for many quantum spectroscopy applications, with shorter correlation times typically yielding better performance. Furthermore, it provides fundamental insight into the effects of dispersion on the bi-photon state. Here, we retrieve ultrashort bi-photon correlation times of around $100\,\mathrm{fs}$ by measuring simultaneously spectral and temporal interferograms at the output of an SU(1,1) interferometer based on an integrated broadband parametric down-conversion source in a $\mathrm{Ti:LiNbO}_3$ waveguide.
\end{abstract}

\maketitle

%\tableofcontents

\section{Introduction}

Quantum spectroscopy has raised significant interest as a tool for the investigation of materials and light sensitive biological samples at the single photon level \cite{Mukamel2020, Brecht2015, Shaul2016, Dayan2004}. In classical spectroscopy, linear interferometers are widely used to estimate the phase properties of a system under test, e.g., in spectral phase interferometry, direct field reconstruction or linear interference \cite{Kane1993, Iaconis1998}. An alternative system are non-linear interferometers, also called SU(1,1) interferometers, that contain active optical elements and thereby provide different mechanisms of phase detection compared to linear interferometers \cite{Yurke1986, Chekhova2016, Li2014}. These interferometers have attracted much attention as it is possible to achieve superior phase sensitivity with them \cite{Manceau2017, Santandrea2023, Florez2022}. In addition, SU(1,1) interferometry allows for operating the different interferometric paths with different colours. This has enabled various protocols of quantum metrology, such as spectroscopy, imaging or optical coherence tomography with undetected photons; these are of particular interest for life science experiments \cite{Lemos2014, Lindner2021, Toepfer2022, Paterova2018, Kaufmann2022, Kviatkovsky2020, Rojas-Santana2021, Valles2018} because they separate the detected wavelength from the probe light and remove the need for detecting light at unfavourable wavelengths such as in the mid infrared (MIR).
The realization of these measurement concepts raises the need for bright and broadband bi-photon sources that generate strong time-frequency entanglement. A high source brightness is typically achieved with integrated optics solutions that offer long interaction lengths and tight field confinement. Broadband operation then requires a careful engineering of the waveguide dispersion \cite{Uren2009}. Strong time-frequency entanglement, finally, is characterized by the product of the bi-photon correlation bandwidth - in typical cases and in this paper the linewidth of a continuous wave pump laser - and the bi-photon correlation time. The latter generally limits the performance of quantum spectroscopy by limiting achievable resolutions. Furthermore the correlation time provides fundamental insights into the effects of dispersion on the broadband bi-photon \cite{Baek2008, Baek2008_2,Valencia2002, Harris2007}.\\
As the correlation times of broadband sources typically lie in the order of $10-100\,\mathrm{fs}$, they are hard to measure due to insufficient detector timing resolution. Classical characterization techniques for ultra-short pulses, e.g., FROG or SPIDER, rely on nonlinear interactions that require high intensities and are therefore not applicable to single photons \cite{Kane1993, Iaconis1998}. Therefore, a new scheme for measuring the correlation time of single photon sources is required.

Here we show that the correlation time of a broadband, integrated bi-photon source can be measured by incorporating the source into an SU(1,1)-interferometer. 
We generate bi-photons via parametric down-conversion (PDC) in a dispersion-engineered, periodically poled titanium-indiffused waveguide in lithium niobate. We note that the concepts in this work equally apply to photon pair sources based on four-wave mixing, although we focus on PDC.
In the experiment, we observe a characteristic dependence of the width of the temporal interferograms for varying amounts of second order dispersion inside the interferometer. 
Furthermore, we show that the observed characteristic dependence is affected by the spectral shape of the PDC source. We develop a supporting theory, which allows us to find a functional dependence between temporal and spectral measurements and to extract the Fourier-limited correlation time. This information is key for further applications, e.g., entangled two-photon absorption where knowledge about the bi-photon correlation time is a crucial parameter for correctly retrieving the absorption cross-section.

\section{Concept}
\label{sec:concept}

\begin{figure*}[ht!]
    \centering
    \includegraphics[trim={0.5cm 18cm 1cm 1cm}, clip, width = \textwidth]{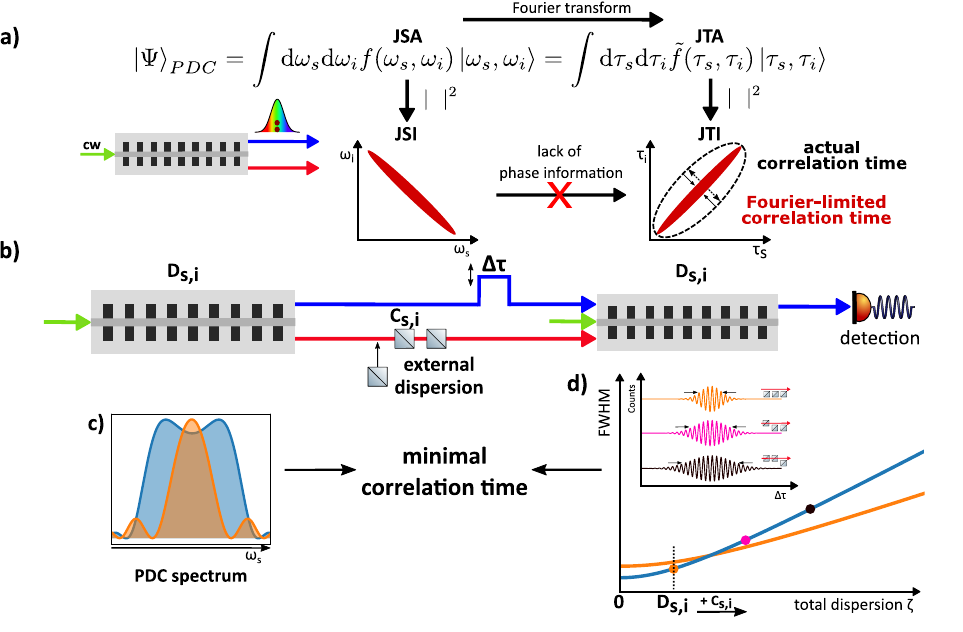}
    \caption{a) For a given PDC state of a broadband bi-photon source, described by the JSA in the spectral domain and the JTA in the temporal domain, the directly measurable JTI does not allow to retrieve the JTI due to a lack of phase information. The JTI might therefore be broadened due to second order dispersion. b) Incorporating a broadband PDC source in an SU(1,1) interferometer allows one to extract the correlation time for different PDC spectra by measuring the characteristic dependence of the temporal interference width for different amounts of total second order dispersion in the interferometer. This contains the intrinsic waveguide dispersion for the signal and idler photons, $D_{s,i}$, and an additional external dispersion, $C_{s,i}$.}
    \label{fig:concept}
\end{figure*}

The bi-photon state from the photon pair generation process in a nonlinear material can be written as $|\psi\rangle_\mathrm{PDC} = \int \, \mathrm{d}\omega_\mathrm{s} \mathrm{d}\omega_\mathrm{i} f(\omega_\mathrm{s}, \omega_\mathrm{i}) |\omega_\mathrm{s},\omega_\mathrm{i}\rangle = \int \, \mathrm{d}\tau_\mathrm{s} \mathrm{d}\tau_\mathrm{i} \tilde{f}(\tau_\mathrm{s},\tau_\mathrm{i}) |\tau_\mathrm{s},\tau_\mathrm{i}\rangle$. It is described via the joint spectral amplitude (JSA) $f(\omega_s,\omega_i)$ in the frequency domain, which contains information about time-frequency entanglement between signal and idler photons $\omega_{s,i}$ in a generated state. In the temporal domain, the joint temporal amplitude (JTA) $\tilde{f}(\tau_s,\tau_i)$ is connected to the JSA via Fourier transform and contains information about the correlation time as the uncertainty in signal and idler arrival times $\tau_{s,i}$ \cite{Grice1997, Law2000}. The joint spectral intensity (JSI), the modulus square of the JSA, of a broadband PDC can be directly measured with established methods \cite{Avenhaus2009}. However, the joint temporal intensity (JTI), the modulus square of the JTA, and thereby the bi-photon correlation time, defined as the FWHM of the JTI \cite{Uren2009}, is hard to measure directly \cite{Kuzucu2008}. Furthermore, it cannot generally be retrieved from a Fourier transform of the JSI; a possible complex phase on the JSA whose information gets lost when measuring the JSI can lead to an increased correlation time and associated broadening of the JTA and thus, consequently, the JTI, as schematically shown in Fig. \,\ref{fig:concept} a).\\

In our measurement concept, shown in Fig.\,\ref{fig:concept} b), we build an SU(1,1) interferometer from two identical bi-photon sources\textemdash in our case a PDC source in a double pass configuration detailed later\textemdash and record temporal interferograms observed by counting signal and idler photons at the output of the interferometer when introducing a relative temporal delay between the two photons. We show that the correlation time of the bi-photon can be determined by monitoring the width of these interferograms as the amount of second order dispersion in the interferometer is varied as depicted in the inset of Fig.\,\ref{fig:concept} d).

By varying the spectral envelope of the generated PDC shown in Fig.\,\ref{fig:concept} c) and observing the FWHM of the temporal interferogram, we notice different characteristic dependencies. From this behaviour we can extract the minimal that is, the Fourier limited correlation time of the bi-photon as well as the actual correlation time that is subject to the material dispersion of the source for different source operating conditions, see Fig.\,\ref{fig:concept} d). 

Our method is based on a model of an SU(1,1) interferometer including a broadband PDC process. For this, we start from the JSA of a single PDC waveguide pumped by a cw-laser, which is given by \cite{Grice1997}:

\begin{equation}
    f(\Delta \omega)  = \mathrm{sinc} \left(\frac{\Delta \beta(\Delta\omega)L}{2} \right) e^{i\cdot \frac{\Delta \beta(\Delta \omega)L}{2}}
\label{eq:jsa}
\end{equation}

where $\Delta \omega = \omega_s - \bar{\omega}_s = - (\omega_i-\bar{\omega}_i)$ is the frequency detuning from the central signal frequency $\bar{\omega}_s$, $L$ the length of the waveguide and $\Delta \beta$ the phase mismatch between the pump and the generated signal and idler fields. The exponential term in this expression contains the phase caused by the dispersion of the waveguide.

Achieving the desired short correlation times requires the production of a broad bi-photon spectrum. Long waveguides typically generate narrow bi-photon spectra, especially in the case of two colour sources. Therefore, we are using dispersion engineering to achieve a PDC source where signal and idler group velocities are matched to generate broader spectra. If we now expand $\Delta \beta$ up to second order \cite{Uren2005} and make use of the strong frequency correlations due to cw-pumping, we arrive at a phase-mismatch $\Delta \beta$ that only shows quadratic terms in $\Delta \omega$ to leading order, as presented in Appendix \ref{sec:delta_beta}. With this, the JSA can be written as
\begin{equation}
\begin{aligned}
f(\Delta \omega)  = \mathrm{sinc} \left(\Delta\gamma \cdot \Delta\omega^2\right) e^{i\cdot (\Delta \gamma \cdot \Delta \omega^2)}.
\label{eq:JSA_approx}
\end{aligned}    
\end{equation}
It becomes apparent that the group delay dispersion associated to the waveguide $\Delta \gamma = - \frac{(D_s + D_i)L}{2}$, including the second order dispersion of signal $D_s$ and idler $D_i$, defines the bandwidth of the JSA in this approximation.
Combining two of our PDC sources in the interferometer and including a phase between them yields a total joint spectral amplitude of the whole interferometer $f_{SU(1,1)}(\Delta \omega)$. This expression includes a phase caused by the intrinsic second order dispersion from the waveguide $(D_s + D_i)L \Delta \omega^2$ and an externally applied phase $\Phi_{ext}(\Delta \omega)$, cf. Appendix \ref{sec:JSA}:

\begin{equation}
    f_{SU(1,1)}(\Delta \omega) = f(\Delta \omega)\cdot \left( 1 + e^{i[(D_s + D_i) L \Delta \omega^2  - \Phi_{ext}(\Delta \omega)]}\right)
\label{eq:su11_jsa}
\end{equation}

We consider that the experimentally relevant external phase, $\Phi_{ext}(\Delta \omega)$, consists only of a linear and a quadratic phase. A linear phase can be introduced by a time delay in either one or both of the arms $\Delta t_{s,i}$. The external quadratic phase is realised by placing a dispersive material in the signal or idler arm, which, in addition to a time delay, imprints a quadratic phase characterized by the chirp parameters $C_{s,i}$. The resulting JSA of the full system is presented in Appendix \ref{sec:JSA_phase}. The counts in the signal arm as a function of the delay $\Delta t_s$ and external second order dispersion in signal and idler arm $C_{s,i}$ can be calculated by integrating over the detuning $\Delta \omega$:

\begin{equation}
\begin{aligned}
    &S(\Delta t_s,\Psi) = \int_{-\infty}^{\infty} \mathrm{d}\Delta \omega \bigg|f_{SU(1,1)}(\Delta \omega)\bigg|^2\\
    &=\int_{-\infty}^{\infty} \mathrm{d}\Delta \omega \,\mathrm{sinc}^2(\Delta \gamma \Delta \omega^2)\cdot \mathrm{cos}^2(\Psi).
\end{aligned}
\label{eq:integral}
\end{equation}
Here, the external second order dispersion together with the intrinsic waveguide dispersion result in a total phase $\Psi =\frac{1}{2} (\Phi_{ext} - (D_s + D_i)L \Delta \omega^2)$.
Eq.\,\ref{eq:integral} is the final expression that is used to calculate the temporal interferograms presented in this work via numerical integration.\\
To aid in understanding, an approximate analytical solution can be derived by approximating the $\mathrm{sinc}^2$ term as a Gaussian, as detailed in Appendix \ref{sec:approx_sol}:

\begin{equation}
\begin{aligned}
    & S(\Delta t_s, C_s,C_i) \approx \\ & \frac{1}{2} + A \cdot \mathrm{exp}\left(\frac{\Delta t^2_s}{\sigma_{env}^2}\right)  \cdot \mathrm{cos}\left(\frac{\zeta \Delta t^2_s}{4 \zeta^2+\Delta \gamma^2}-\bar{\omega}_s \Delta t_s\right)
\label{eq:fringing}
\end{aligned}
\end{equation}

\noindent The width of the Gaussian envelope of the temporal interference pattern $\sigma_{env}^2$ is given by

\begin{equation}
\begin{aligned}
    \sigma_{env}^2 = \frac{2 \zeta}{\Delta \gamma}+ 0.5\Delta \gamma,
\label{eq:envelope}
\end{aligned}
\end{equation}

\noindent where $\zeta = \Delta \gamma - 0.5(C_s+C_i)$ is the total second order dispersion in the system. 
From this expression we can see that the minimal width of the temporal interferogram, i.e. without any second order dispersion, is given by $0.5 \Delta \gamma$ and thereby dependent on the bandwidth of the JSA, as shown in Appendix \ref{sec:approx_sol}. This situation can be reached by cancellation of the intrinsic waveguide dispersion by anomalous external dispersion. For increasing amounts of external chirp $C_{s,i}$, the width of the temporal interferogram increases faster for smaller $\Delta \gamma$. This is because external second order dispersion leads to a quicker broadening on shorter pulses, a behaviour known in classical ultra-fast optics. This leads to a distinctive behaviour for different spectral envelopes of the underlying JSA and thereby correlation times, as schematically shown in Fig.\,\ref{fig:concept} d). This property has been exploited in the following experimental scheme to find the characteristic dependence between the temporal interference width and the second order dispersion, where we find that not only the spectral bandwidth, but also the spectral shape influences the temporal correlations.

\section{Experimental Setup}
\label{sec:setup}
The core of the interferometer is a periodically poled Ti:LiNbO$_3$ guided-wave PDC source with matching signal and idler group-velocities, a design that is similarly used in the quantum pulse gate \cite{Eckstein2011}. The use of a waveguide as PDC source generally leads to strong frequency correlations in the JSA while spatial correlations between the photons do not arise due to generation in single spatial modes. This waveguide is pumped by a $514\,\mathrm{nm}$ cw-laser, generating PDC light at $830\,\mathrm{nm}$ and $1360\,\mathrm{nm}$ in signal and idler, respectively. We can reach a spectral bandwidth of more than $6\,\mathrm{THz}$ (FWHM) with our $40\,\mathrm{mm}$ long waveguide. 
The input facet of the waveguide is coated with an anti-reflection coating for the pump wavelength, while the output facet features a high-reflectivity coating for the pump light to reflect the forward-propagating pump into the reverse direction in the same spatial mode. Both facets are coated with anti-reflection coatings for the signal and idler wavelengths.

\begin{figure}[ht!]
    \centering
    \includegraphics[trim={0cm 1cm 0cm 0.5cm},clip,width=0.7\linewidth]{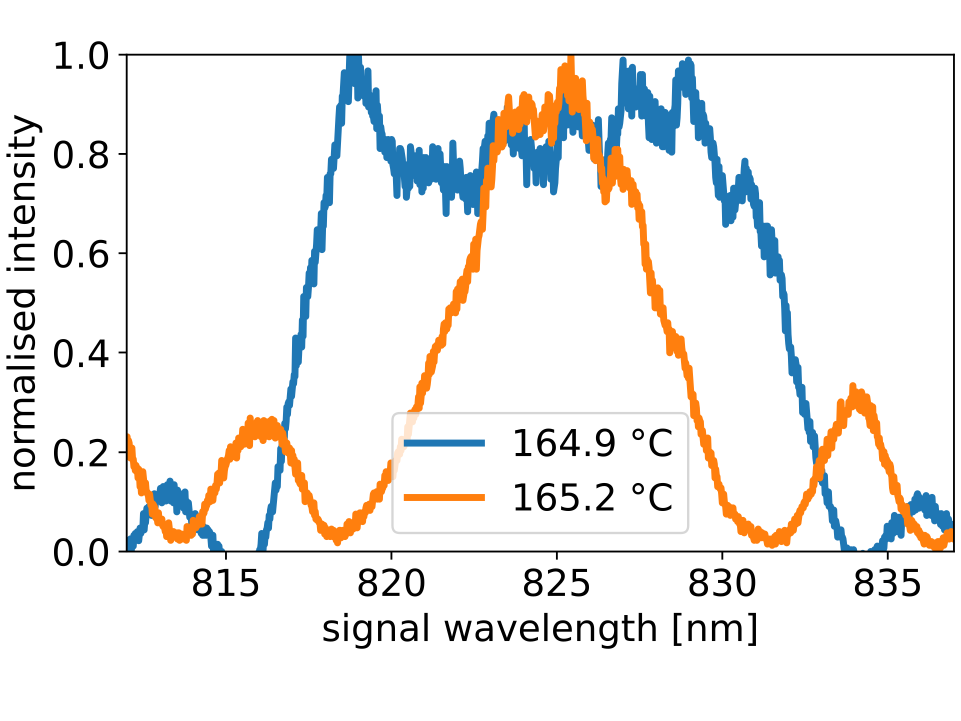}
    \caption{Measured PDC spectra for temperatures of $164.9\,^\circ \mathrm{C}$ (blue) and $165.2\,^\circ \mathrm{C}$ (orange) with spectral bandwidths of $6.46\,\mathrm{THz}$ and $2.82\,\mathrm{THz}$. The data is normalised to the maximum intensity of each spectrum and the background of the spectrometer is removed.}
    \label{fig:spectra}
\end{figure}

The shape and bandwidth of the emitted PDC spectrum can be tuned by adjusting the waveguide temperature. Different spectral bandwidths then result in different correlation times of the bi-photon. We operated our source in a regime with two different spectra as shown in Fig.\,\ref{fig:spectra}, c.f. \cite{Pollmann2023}. They cover a spectral bandwidth of $6.46\,\mathrm{THz}$ (nominal operating point of the source, blue curve) and $2.82\,\mathrm{THz}$ (detuned temperature, orange curve). The temperature difference of the waveguide between these spectra is $0.3\,\mathrm{K}$. Due to the transition from a quasi top-hat to a $\mathrm{sinc}^2$ spectral shape, the presented spectra have different time-bandwidth products (TBP). We calculate them from the FWHM of the simulated spectral and temporal intensities to be around 0.5 and 0.2, respectively, which will be important for the interpretation of our results.

\begin{figure}[ht!]
    \centering
    \includegraphics[trim={0 13.5cm 0 5.5cm}, clip,width=\linewidth]{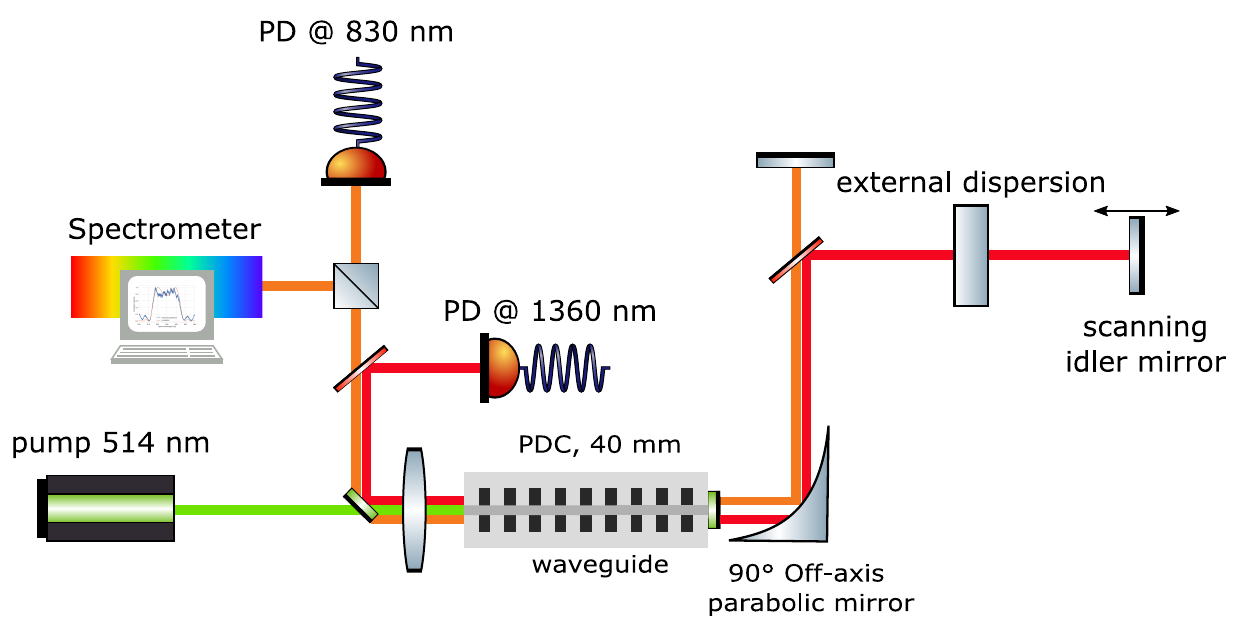}
    \caption{The experimental setup consisting of a 40 mm waveguide as a source for PDC photons in a folded Mach-Zehnder-geometry. The setup allows to add external dispersion in either of the two arms of the interferometer. The end mirror of the idler arm can be moved by a motorized stage. An off-axis parabolic mirror collimates the signal and idler beams and focuses them back into the waveguide. After the second interaction, signal and idler photons are separated from the pump and detected by avalanche photo-diodes or a spectrometer for the signal case.}
    \label{fig:setup}
\end{figure}

A schematic of the experimental setup is shown in Fig.\,\ref{fig:setup}. There, signal and idler photons are generated in the first pass of the pump light through the waveguide and collimated by an off-axis parabolic mirror (OAP). We separate the signal and idler photons using a dichroic mirror. The end mirror of the idler arm can be moved by a motorized stage (PI M111.1DG) to introduce a delay between both arms, resulting in a linear spectral phase. Different amounts of dispersive optical materials can be introduced in one or both of the interferometer arms to imprint a quadratic phase on the bi-photon. After reflection on the end mirrors, we couple the signal and idler photons back into the same waveguide via the OAP. The presented design ensures high spectral interference visibility, because the two PDC sources are identical by design.

The fields in the reverse direction are again separated using dichroic mirrors and are detected with a Si-APD (Perkin Elmer Photon Counting Module SPCM-AQR-14FC) for the signal light around $830\,\mathrm{nm}$ and an InGaAs-APD (IDQ id200) for the idler light around $1360\,\mathrm{nm}$ for photon counting.

The data from the avalanche photo-diodes has been recorded with a Swabian Instruments Time-Tagger 20, as the idler mirror was scanned. For each step, the counts have been recorded for an integration time of $0.3\,\mathrm{s}$. Without altering the setup, the spectrum of the signal field can be recorded on a single photon sensitive spectrometer (Andor Shamrock SR-500i spectrograph with Newton 970P EMCCD - camera) for a fixed position of the idler mirror. Each spectrum has been acquired with an integration time of $1\,\mathrm{s}$ and a spectral resolution of about $30\,\mathrm{GHz}$.

\section{Results}
\label{sec:results}
The total second order dispersion is determined by recording spectra of the signal photons for a fixed idler mirror position resulting in a difference in the interferometers arm lengths of about $0.3\,\mathrm{mm}$. This allows optimal phase extraction by achieving a balance between maximizing the number of visible interference fringes and ensuring that the measurement is not limited by the resolution of the spectrometer.

\begin{figure}
    \centering
    \includegraphics[trim={0cm 0cm 0cm 0cm}, clip, width=\linewidth]{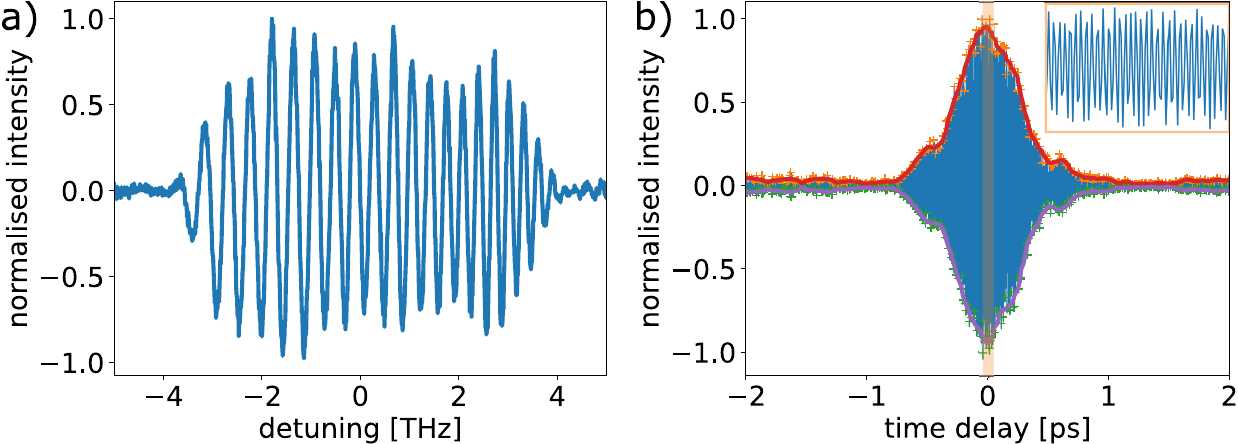}
    \caption{a) Spectral interferogram used to extract the second order dispersion in the system. b) The corresponding temporal interferogram with upper and lower envelope, which are used to obtain the temporal interference width. The inset shows a zoom in the marked region.}
    \label{fig:temp_spec}
\end{figure}

The second order dispersion can be estimated by fitting a cosine function to the spectral interferogram. The spectral intensity $S(\Delta \nu)$ with a frequency detuning $\Delta \nu$ around the central frequency is given by \cite{Riazi2019}: 

\begin{equation}
    S(\Delta \nu) \propto f(\Delta \nu)^2(1 + \cos{\left(\Psi(\Delta \nu)\right)}) .
    \label{eq:spec_int}
\end{equation}

A typical spectral interferogram is shown in Fig.\,\ref{fig:temp_spec} a). This interferogram has been retrieved from the raw data by subtracting a reference spectrum and normalising the resulting data to their maximum value. A region of $4\,\mathrm{THz}$ detuning around the central frequency is used to fit Eq.\,\ref{eq:spec_int} to ensure a sufficient signal-to-noise ratio for the fit. The uncertainty for the second order dispersion is given by the fit parameters. We arrive at an intrinsic second order dispersion for the waveguide itself of about $|\Delta \gamma| = 14800 \pm 240 \,\mathrm{fs}^2$. This value is close to the theoretical value of $10893\,\mathrm{fs}^2$ calculated from the material dispersion of the waveguide. The observed discrepancy can arise from the dielectric end-facet coatings on the waveguide or other elements in the setup, e.g., dichroic mirrors, as well as from imprecisions in the sellmeier model for the refractive index.

Temporal interferograms were recorded with the same amounts of second order dispersion added to the interferometer as in the spectral measurements for both PDC spectra. To record the interferogram, the idler mirror has been moved over a range of $0.8\,\mathrm{mm}$ in steps of $200\,\mathrm{nm}$ while recording photon counts at the APDs for signal and idler photons at each step. To extract the FWHM of the temporal interferogram, the mean number of counts was subtracted from the raw data and the signal was normalised to its maximum value. Furthermore, we fit and smooth the upper and lower envelope by applying a Savitzky-Golay filter and retrieve the FWHM from the resulting smoothed envelope function, cf. Fig.\,\ref{fig:temp_spec} b).
The error on the FWHM of $11.8\,\%$ is found by measuring the variability within 20 subsequent interferogram measurements with no external dispersion. The resulting relative uncertainty was applied to the other data points.

\begin{figure}[h!]
    \centering
    \includegraphics[trim={2cm 18cm 2cm 3cm}, clip, width=\linewidth]{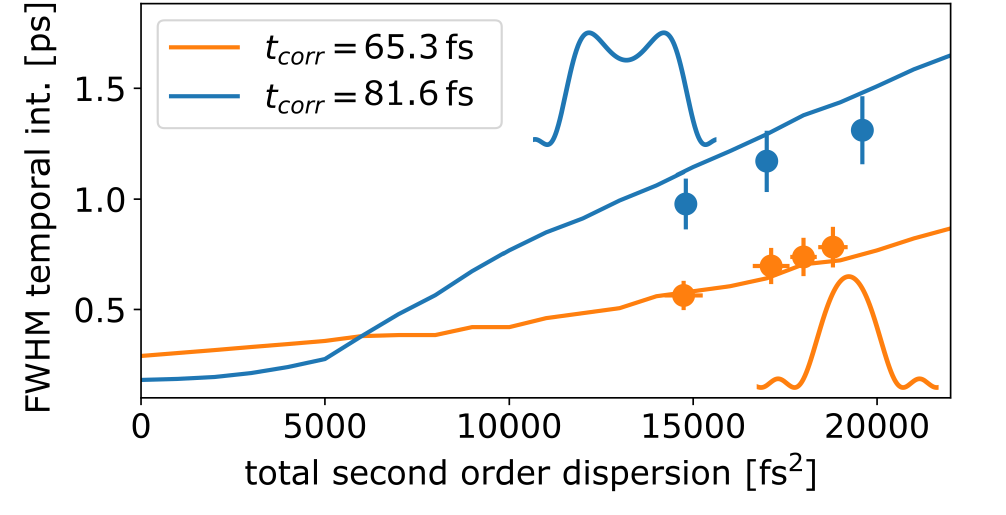}
    \caption{The change in the temporal interference width with varying second order dispersion is shown for two different PDC spectra that are depicted in the corresponding colour in the figure. The solid lines show simulations based on these spectra.}
    \label{fig:fringing}
\end{figure}

The characteristic dependence of the temporal interference width on the total second order dispersion in the system is revealed by combining temporal and spectral measurements as shown in Fig.\,\ref{fig:fringing}. Here, the data points are presented together with simulations of the temporal interference width based on the underlying PDC spectra via Eq.\,\ref{eq:integral}. The spectra are shown schematically in the same colour as the corresponding simulation. The simulations are in good agreement with the measured data and reveal a functional dependence between temporal interference width and second order dispersion for different shapes of the PDC spectra, as can be seen in Eq.\,\ref{eq:envelope}. This allows one to infer the second order dispersion from the temporal interferogram with no need for single-photon sensitive spectrometers. The obtained phase information can then be used to calculate the actual correlation time of the generated bi-photon by including this phase in the Fourier transform of the JSA.

With this information, the actual correlation times at the output of the PDC source, given as the FWHM of the JTI, can be calculated to be $73.5\,\mathrm{fs}$ and $177.5\,\mathrm{fs}$ for the narrower and broader spectrum. We can also calculate the minimal correlation time of bi-photons via Fourier transform of the two PDC spectra, as stated in section II assuming a flat phase profile on the JSA. This results in minimal correlation times of $65.3\,\mathrm{fs}$ for the narrower spectrum, and $81.6\,\mathrm{fs}$ for the broader spectrum. These minimal correlation times can be reached by introducing the correct amount of anomalous second order dispersion in the setup. Note however, that these correlation times vary in a counter-intuitive way as the bandwidths vary drastically. This can be attributed to the different TBP of the different spectral shapes, leading to a shorter correlation time for the narrower spectrum. Furthermore, the exact shape of the temporal interferogram strongly depends on the temporal shape of the bi-photon. In our case, the $1/e^2$ widths of the Fourier-limited JTIs are $191.8\,\mathrm{fs}$ for the narrower spectrum and $171.4\mathrm{fs}$ for the broader spectrum. This is reflected by the behaviour of the FWHM of the temporal interferogram in the case of no second-order dispersion, which is larger for the narrower spectrum, see Fig.\,\ref{fig:fringing}.

\section{Conclusion and Outlook}
\label{sec:conclusion}
We have presented a scheme for determining correlation times for a broadband PDC source utilizing an SU(1,1) interferometer. By performing simultaneous spectral and temporal measurements, we reveal the connection between the second order dispersion in the system and the FWHM of the temporal interferogram. The data agrees with our model for a broadband SU(1,1) interferometer, where the PDC spectrum is the only free parameter. After validation by the characteristic dependence of the width of the temporal interferogram on second order dispersion, the minimal, Fourier-limited, correlation time could be calculated. For two different underlying PDC spectra, correlation times of $65.3\,\mathrm{fs}$ and $81.6\,\mathrm{fs}$ have been extracted.

The obtained knowledge about the correlation time of the bi-photon is important for further experiments, such as entangled two-photon absorption, where the absorption cross section depends critically on the correlation time. Furthermore, the validation of our model enables us to estimate the second order dispersion of an object under test in the long wavelength arm by measuring temporal interferograms in the short wavelength arm, i.e. with undetected photons. Finally, the established connection between spectral and temporal interferograms can be further exploited for source characterization or quantum state manipulation.

\begin{acknowledgements}
F.R. is part of the Max Planck School of Photonics supported by the German Federal Ministry of Education and Research (BMBF), the Max Planck Society, and the Fraunhofer Society. 
We acknowledge financial support from the Federal Ministry of Education and Research (BMBF) via the grant agreement no. 13N15065 (MiLiQuant). 
This project has received funding from the European Union’s Horizon Europe research and innovation programme under grant agreement No 101070700 (MIRAQLS). 
\end{acknowledgements}

\appendix
\section{Taylor expansion phase mismatch}
\label{sec:taylor}
We start by performing the Taylor expansion of the phase-mismatch $\Delta \beta$ for the specific case of pumping the PDC with a cw laser. The full expression, already written in terms of the frequency detuning $\Delta \omega = \omega_s - \bar{\omega}_s = - (\omega_i-\bar{\omega}_i)$, takes the form:

\begin{equation}
\begin{aligned}
    &\Delta \beta(\omega_s,\omega_i) = k_p(\omega_s+\omega_i) -k_s(\omega_s)-k_i(\omega_i)\\
    &= \Delta \beta^{(0)} + (\kappa_s - \kappa_i) \Delta \omega + (\eta_s + \eta_i) \Delta \omega^2 + \eta_p \Delta \omega \cdot (-\Delta \omega) 
\end{aligned}
\end{equation}

Here, we assume that $\Delta \beta^{(0)}$ can be set to zero by periodic poling. To obtain the negative sign in the last term, we made use of the fact, that $\Delta \omega = (\omega_s - \bar{\omega}_s) = - (\omega_i - \bar{\omega}_i)$ due to the strong frequency correlations in cw pumping. In the above expression, $\kappa_s = L\left(\frac{\partial k_p}{\partial \omega}\bigg|_{\omega_s+\omega_i}-\frac{\partial k_s}{\partial \omega}\bigg|_{\omega_s}\right)$ and analogous for $k_i$, $\eta_s = \frac{L}{2}\left(\frac{\partial^2 k_p}{\partial \omega^2}\bigg|_{\omega_s+\omega_i}- \frac{\partial^2 k_s}{\partial \omega^2}\bigg|_{\omega_s}\right)$ and analogous for $k_i$, lastly $\eta_p = L \frac{\partial^2 k_p}{\partial \omega^2}\bigg|_{\omega_s+\omega_i}$. We see that terms containing $\frac{\partial k_p}{\partial \omega}$ and $\frac{\partial^2 k_p}{\partial \omega^2}$ are canceled out as we consider the specific case of cw pumping.\\

\section{Phase-mismatch for group velocity matched PDC source}
\label{sec:delta_beta}

We now consider the expansion of $\Delta \beta$ in the case of cw-pumping until second order:

\begin{equation}
\begin{aligned}
\Delta \beta(\Delta \omega) \approx & - \frac{\partial k_s}{\partial \omega}\bigg|_{\bar{\omega}_s} \Delta \omega + \frac{\partial k_i}{\partial \omega}\bigg|_{\bar{\omega}_i} \Delta \omega \\
& - \frac{\partial^2 k_s}{\partial \omega^2}\bigg|_{\bar{\omega}_s} \Delta \omega^2 - \frac{\partial^2 k_i}{\partial \omega^2}\bigg|_{\bar{\omega}_i} \Delta \omega^2,
\end{aligned}
\end{equation}

where $\frac{\partial k_s}{\partial \omega}\bigg|_{\bar{\omega}_s} = 1/v_{g,s}$ and $\frac{\partial k_i}{\partial \omega}\bigg|_{\bar{\omega}_i} = 1/v_{g,i}$ are the inverse signal and idler group velocities, respectively. The second order derivatives are connected to the group velocity dispersion of signal and idler fields by $D_{s,i} = \frac{\partial^2 k_{s,i}}{\partial \omega^2}\bigg|_{\bar{\omega}_{s,i}}$.\\
From this derivation we can see that a broad PDC spectrum can be achieved by group-velocity matching, which requires tailoring the dispersion in a way that $v_{g,s} = v_{g,i}$. This leads to a phase-mismatch $\Delta \beta$ that now only shows a quadratic term to leading order:

\begin{equation}
    \Delta \beta(\omega_s,\omega_i) \approx - \frac{\partial^2 k_s}{\partial \omega^2}\bigg|_{\bar{\omega}_s} \Delta \omega^2 - \frac{\partial^2 k_i}{\partial \omega^2}\bigg|_{\bar{\omega}_i} \Delta \omega^2.
\end{equation}

\section{Influence of external phase}

An external phase, e.g. second order dispersion, on the joint spectral amplitude can be described by \cite{Uren2009}:

\begin{equation}
    f(\omega_s,\omega_i)= \Phi(\omega_s,\omega_i)\alpha(\omega_s+\omega_i)E(\omega_s,\omega_i),
\end{equation}

where $\Phi(\omega_s,\omega_i)$ is the phase matching function, $\alpha(\omega_s+\omega_i) = \delta(\omega_s+\omega_i)$ is the pump function for cw pumping and $E(\omega_s,\omega_i)=\mathrm{exp}[iC_s\Delta \omega^2]\mathrm{exp}[iC_i\Delta \omega^2]$ is the introduced external dispersion with chirp parameters $C_s$ and $C_i$ in signal and idler path, respectively.

\section{SU(1,1) joint spectral amplitude}
\label{sec:JSA}
To construct the joint spectral amplitude at the output of the SU(1,1) interferometer, i.e. after going through the second PDC process, we consider the two individual amplitudes first and sum them up afterwards.\\
The photons produced in the first pass of the pump field through the PDC source experience the second order dispersion introduced in either of the interferometer arms as well as dispersion from the waveguide material itself, during the generation process and when passing through the waveguide for the second time. We are assuming that no nonlinear process occurs during that second pass, such that the generated photons only propagate through the waveguide, but do not interact with the pump field. We can write the JSA as:

\begin{equation}
\begin{aligned} 
    f(\Delta \omega)_1=& f(\Delta \omega) \cdot \mathrm{exp}[-i(D_s L + C_s) \Delta \omega^2]\\ & \cdot \mathrm{exp}[-i(D_i L + C_I) \Delta \omega^2]
\end{aligned}
\end{equation}

For the second pass of the pump field, the photons do not propagate through the interferometer arms and thus only acquire the phase included in the fundamental JSA $f_2(\Delta \omega)=f(\Delta \omega)$ already, given in equation \ref{eq:jsa}.\\
This results in the total JSA given by:

\begin{equation}    
\begin{aligned}
    f_{SU(1,1)}(\Delta \omega)= &f_1(\Delta \omega)+f_2(\Delta \omega)\\
    & =f(\Delta \omega) (1+ e^{-i(D_s L +D_i L+C_s+C_i) \Delta \omega^2})
\end{aligned}
\end{equation}

It can be seen directly from this expression that, due to the lack of cross-terms, the quadratic phase in either of the arms can be compensated by placing the corresponding amount of dispersion in either the same or the other arm of the interferometer.

\section{JSA for internal and external phase}
\label{sec:JSA_phase}

The external phase applied between the two PDC sources can be written as:

\begin{equation}
\begin{aligned}
    \Phi_{ext}(\Delta \omega) &= \omega_s \Delta t_s + C_s \Delta \omega^2 + C_i \Delta \omega^2 \\
     & =  (\bar{\omega}_s + \Delta \omega) \Delta t_s + C_s \Delta \omega^2 + C_i \Delta \omega^2
\end{aligned}
\end{equation}

We now consider varying $\Delta t_s$, $C_s$ and $C_i$ and write the total phase $\Psi$ in the exponent of equation \ref{eq:su11_jsa} as:

\begin{equation}
    \Psi = \frac{1}{2} \big([(D_s+D_i)L-C_s-C_i]\Delta \omega^2 - \Delta t_s \bar{\omega}_s - \Delta t_s \Delta \omega \big)
\end{equation}

 When expressing the real part of the exponential function as a cosine function, the total JSA becomes:

\begin{equation}
\begin{aligned}
    f_{SU(1,1)}(\Delta \omega,C_s,C_i,\Delta t_s) = & 2 f(\Delta \omega)\cdot \cos(\Psi) \cdot \mathrm{exp}(i\frac{\Psi}{2})
\end{aligned}
\end{equation}

\section{Analytical approximate solution}
\label{sec:approx_sol}

To find an approximate analytical solution, we split up the cosine term in Eq.\,\ref{eq:integral} into three terms:

\begin{equation}
    \mathrm{cos}^2(\Psi)= \frac{1}{2}+ \frac{1}{4}\mathrm{e}^{2i\Psi}+ \frac{1}{4}\mathrm{e}^{-2i\Psi}, 
\end{equation}
such that $S(\Delta t_s, C_s,C_i) = I_1 + I_2 + I_3$.
Furthermore, the $\mathrm{sinc}^2$ term can be approximated by a Gaussian function in the latter two integrals:
\begin{equation}
    \mathrm{sinc}^2(\Delta \gamma \Delta \omega^2) \approx \mathrm{exp}\left(-\frac{\Delta \gamma}{\sigma}\Delta \omega^2\right)
\label{eq:sinc}
\end{equation}
with $\Delta \gamma = -\frac{(D_s+D_i)L}{2}$ and $\sigma = 2$ to match the FWHM of both functions.
The first integral yields a constant offset of $0.5$, as this is a know integral of a Gaussian. With the approximation $I_2$ becomes:

\begin{equation}
\begin{aligned}
    &I_2 = A \cdot \mathrm{exp}\left(-i \bar{\omega}_s\Delta t_s\right) \\ & \cdot \mathrm{exp}\left(\frac{\Delta t^2_s(i\Delta \gamma - 0.5 i (C_s+C_i)+0.5 \Delta \gamma)}{\Delta \gamma + 2(\Delta \gamma - 0.5 (C_s+C_i))^2}\right)
\end{aligned}
\end{equation}
 where $A = \frac{2}{3|\Delta \gamma|^{1/2}\sqrt{-\Delta \gamma +C_s+C_i}}$. $I_3$ yields the complex conjugate of $I_2$. We can identify $\zeta = \Delta \gamma - 0.5 (C_s+C_i)$ as the total second order dispersion in the system. Put together, this results in a fringing pattern given by:

\begin{equation}
\begin{aligned}
    & S(\Delta t_s, C_s,C_i) = \\ & \frac{1}{2} + A \cdot \mathrm{exp}\left(\frac{\Delta \gamma \Delta t^2_s}{2 \zeta^2 + 0.5 \Delta \gamma^2}\right)  \cdot \mathrm{cos}\left(\frac{\zeta \Delta t^2_s}{4 \zeta^2+\Delta \gamma^2}-\bar{\omega}_s \Delta t_s\right)
\label{eq:fringing}
\end{aligned}
\end{equation}
The frequency of the interference fringes is mainly given by the central frequency of the light in the signal arm $\bar{\omega}_s$ but also slightly influenced by the second order phase in the interferometer, due to a the time delay of different frequencies within the pulse caused by dispersion. The pre-factor $A$ determines the visibility of the interference pattern, which also depends on internal and externally applied dispersion. It becomes apparent that the width of the envelope of the temporal interferogram is given by the intrinsic dispersion of the nonlinear material, represented by the parameter $\Delta \gamma$, and can be altered by adding dispersive elements in the signal or idler arm introducing $C_s$ and $C_i$, respectively. It is thereby irrelevant how the second order phase is distributed between the photons, an effect that is also known as non-local dispersion cancellation \cite{Franson1992}.

\bibliographystyle{apsrev4-1}
\bibliography{correlation_time}

\end{document}